\documentclass[10pt,romanappendices,conference,twocolumn]{IEEEtran}

\usepackage{times}
\usepackage{latexsym}
\usepackage{stfloats}
\usepackage{citesort}
\usepackage{graphicx, amsfonts,amsmath, amssymb,array,url,stfloats}
\usepackage[noadjust]{cite}
\usepackage{graphicx}
\usepackage{subfigure}
\usepackage{enumerate}

\newtheorem{corollary}{Corollary}
\newtheorem{proposition}{Proposition}
\newtheorem{lemma}{Lemma}
\newtheorem{definition}{Definition}
\newtheorem{remark}{Remark}

\def\diag{\mathrm{diag}}

\begin{document}

\title{On the MIMO Capacity with Residual \\Transceiver Hardware Impairments}
\author{\IEEEauthorblockN{Xinlin~Zhang\IEEEauthorrefmark{1}, Michail~Matthaiou\IEEEauthorrefmark{2}\IEEEauthorrefmark{1}, Emil~Bj\"ornson\IEEEauthorrefmark{3}\IEEEauthorrefmark{4}, Mikael Coldrey\IEEEauthorrefmark{5}, and M\'erouane~Debbah\IEEEauthorrefmark{3}}
\IEEEauthorblockA{\IEEEauthorrefmark{1}Department of Signals and Systems, Chalmers University of Technology, Gothenburg, Sweden}
\IEEEauthorblockA{\IEEEauthorrefmark{2}School of Electronics, Electrical Engineering and Computer Science, Queen's University Belfast, Belfast, U.K.}
\IEEEauthorblockA{\IEEEauthorrefmark{3}Alcatel-Lucent Chair on Flexible Radio, SUPELEC, Gif-sur-Yvette, France}
\IEEEauthorblockA{\IEEEauthorrefmark{4}Department of Signal Processing, ACCESS, KTH Royal Institute of Technology, Stockholm, Sweden}
\IEEEauthorblockA{\IEEEauthorrefmark{5}Ericsson Research, Ericsson AB, Gothenburg, Sweden}
E-mail: xinlin@chalmers.se, m.matthaiou@qub.ac.uk, emil.bjornson@supelec.fr, \\mikael.coldrey@ericsson.com,  merouane.debbah@supelec.fr}
\maketitle

\begin{abstract}
Radio-frequency (RF) impairments in the transceiver hardware of communication systems (e.g., phase noise (PN), high power amplifier (HPA) nonlinearities, or in-phase/quadrature-phase (I/Q) imbalance) can severely degrade the performance of traditional multiple-input multiple-output (MIMO) systems. Although calibration algorithms can partially compensate these impairments, the remaining distortion still has substantial impact. Despite this, most prior works have not analyzed this type of distortion. In this paper, we investigate the impact of residual transceiver hardware impairments on the MIMO system performance. In particular, we consider a transceiver impairment model, which has been experimentally validated, and derive analytical ergodic capacity expressions for both exact and high signal-to-noise ratios (SNRs). We demonstrate that the capacity saturates in the high-SNR regime, thereby creating a finite capacity ceiling. We also present a linear approximation for the ergodic capacity in the low-SNR regime, and show that impairments have only a second-order impact on the capacity. Furthermore, we analyze the effect of transceiver impairments on large-scale MIMO systems; interestingly, we prove that if one increases the number of antennas at one side only, the capacity behaves similar to the finite-dimensional case. On the contrary, if the number of antennas on both sides increases with a fixed ratio, the capacity ceiling vanishes; thus, impairments cause only a bounded offset in the capacity compared to the ideal transceiver hardware case.

\end{abstract}
\thispagestyle{empty}


\vspace{0.2cm}
\section{Introduction}
MIMO wireless communication systems have attracted considerable attention over the past decades due to their ability to enhance the channel capacity and transmission reliability.  Telatar and Foschini have respectively shown in \cite{telatar1999capacity} and \cite{foschini1998a} that there is a linear growth in channel capacity by increasing the number of transmit and receive antennas, without requiring additional transmit power or bandwidth. Although numerous publications have appeared in this field, the vast majority assumes ideal RF hardware. However, this assumption is quite unrealistic in practice. More specifically, RF impairments, such as I/Q imbalance \cite{qi2010IQ, zou2008IQ}, HPA nonlinearities \cite{qi2012HPA_nonlinearity, au2007HPA} and oscillator PN \cite{durisi2013PN_Wiener, katz2004PN} are known to have a deleterious impact on the performance of practical MIMO systems. Even though one can resort to calibration schemes at the transmitter, or compensation algorithms at the receiver  to partially mitigate these impairments \cite{schenk2008rf}, there still remains certain amount of distortion unaccounted for. The reasons for such residual transceiver impairments are, for example, inaccurate models which are used to characterize the impairments' behavior, imperfect parameters estimation errors due to thermal noise, and unsophisticated compensation algorithms with limited capabilities.

In this context, very few publications have studied the impact of residual transceiver impairments. For example, \cite{studer2010residual} provided experimental results to model the statistical behavior of residual hardware impairments. Moreover, they also investigated the impact of  transmitter impairments on several existing MIMO detection algorithms (e.g., zero-forcing  detection, maximum-likelihood detection, and max-log a posteriori probability detection). In \cite{Emil2013imp}, the authors analyzed the MIMO channel capacity under the aforementioned residual impairment model, but they only considered hardware impairments at the transmitter side and mainly derived high-SNR capacity ceilings. Very recently, \cite{Michalis2013relay} reported how hardware impairments affect dual-hop relaying systems. However, to the best of our knowledge, a detailed study of the MIMO system capacity in the presence of residual transceiver hardware impairments is missing from the literature.


Motivated by the above discussion, we hereafter analytically assess the impact of residual RF impairments in \textit{the transmitter and receiver hardware} of MIMO systems. More specifically, we derive a new analytical expression for the MIMO ergodic capacity in independent and identically distributed (i.i.d.) Rayleigh fading channels for arbitrary SNR values. Additionally, we also present asymptotic capacity expressions in the high-SNR regime. In the low-SNR regime, we derive expressions for the minimum normalized energy per information bit required to convey any positive rate reliably and the wideband slope \cite{lozano2003lowSNR}, which are the two key low-SNR parameters. Throughout our analysis, we find that the impact of residual impairments is marginal on low SNR systems, while it can substantially affect the performance of high SNR systems. In the last part, we analyze the ergodic capacity of large-dimensional MIMO systems with transceiver impairments and deduce asymptotic closed-form expressions for three typical cases. This provides valuable insights on how transceiver impairments affect large-scale (or ``massive'') MIMO systems \cite{Rusek2013a}.



\textit{Notation:} Upper and lower case boldface letters denote matrices and vectors, respectively. The trace of a matrix is expressed by $\mathrm{tr}\left(\cdot\right)$. The $n \times n$ identity matrix is represented by $\mathbf I_n$. The expectation operation is $\mathbb {E} [\cdot]$, while the matrix determinant is denoted by det$(\cdot)$. The superscripts $(\cdot)^ H$ and $(\cdot)^{-1}$ stand for Hermitian transpose and matrix inverse, respectively. The Euclidean vector norm is denoted by $\left\|\cdot\right\|$. The symbol $\mathcal{CN}\left(\mathbf m, \boldsymbol\Sigma\right)$ denotes a circularly-symmetric complex Gaussian distribution with mean $\mathbf m$ and covariance $\boldsymbol\Sigma$.


\section{Signal and System Models}
\label{sec:signalandsystems}

The canonical flat-fading point-to-point MIMO channel with $N_{t}$ transmit antennas and $N_{r}$ receive antennas is modeled as
\begin{equation}
{\mathbf{y}}={\mathbf{H}}\mathbf{s}+\boldsymbol\nu
\label{eq:canonical}
\end{equation}
where ${\mathbf{s}}\in\mathbb{C}^{N_{t}\times 1}$ represents the transmitted signal, with zero mean and covariance matrix $\mathbb{E}_{\mathbf{s}} \left[\mathbf s\mathbf s^H\right] = \mathbf Q$. The received signal is denoted by ${\mathbf{y}}\in\mathbb{C}^{N_{r}\times 1}$, while ${\boldsymbol\nu}  \sim\mathcal{CN}\left(\mathbf{0}, \mathbf{I}_{N_r} \right)$ is the (normalized) additive complex Gaussian receiver noise.
The channel matrix is denoted by $\mathbf{H}\in \mathbb{C}^{N_r\times N_t}$ and is assumed to have i.i.d. complex Gaussian entries with zero mean and unit variance. The receiver is assumed to know $\mathbf{H}$ perfectly, while only its statistical characteristics are available at the transmitter.

\begin{figure}
\begin{centering}
\includegraphics [keepaspectratio,width=\columnwidth]{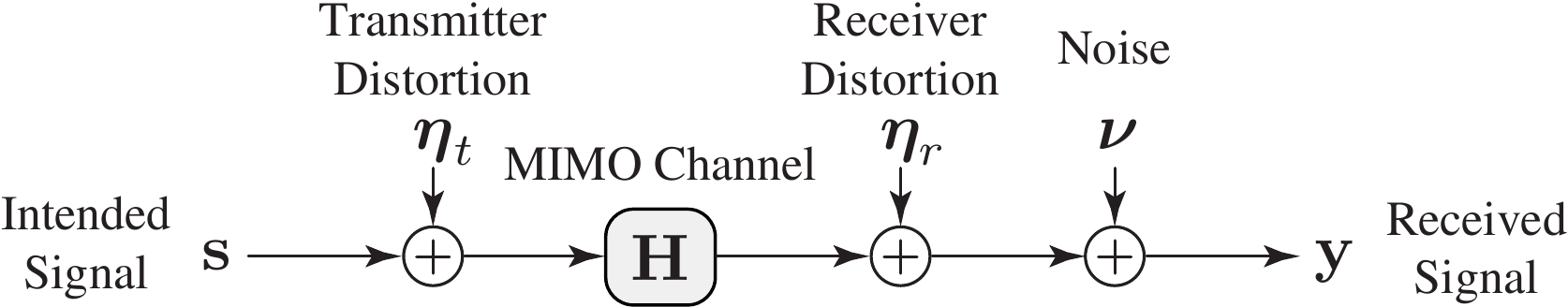}
\caption{Block diagram of the MIMO channel with distortion noises from residual impairments in the transmitter and receiver hardware.}\label{fig:Block_Diagram}
\end{centering}
\end{figure}

Unfortunately, the canonical model cannot describe physical hardware impairments of RF transceivers in an accurate way.
To be more specific, on the transmitter side, the impairments will cause a mismatch between the intended signal and what is actually transmitted; on the receiver side, the impairments will distort the received signal during the reception processing. These impairments come from different sources, for example, I/Q imbalance, HPA non-linearities and PN \cite{schenk2008rf}. Compensation schemes can be applied at both the transmitter and receiver to mitigate part of these impairments; however, as shown in \cite{schenk2008rf,studer2010residual}, the residual impairments will still induce additional additive distortion noises. As proposed and validated in \cite{schenk2008rf,studer2010residual}, the impact of residual transceiver impairments is well-modeled by a more general channel model:
\begin{equation}
{\mathbf{y}} = {\mathbf{H}}\left( {\mathbf{s} + \boldsymbol\eta_{t} } \right) + \boldsymbol\eta_{r} + \boldsymbol\nu
\label{eq:impairmodel}
\end{equation}
where the additive terms $\boldsymbol\eta_t$ and $\boldsymbol\eta_r$ are $\textit{distortion noises}$ from the residual impairments in the transmitter and receiver, respectively. The system block diagram is shown in Fig.~\ref{fig:Block_Diagram}. Furthermore, the measurement results in \cite{studer2010residual} show that the residual transmit distortion noise is well-modeled as Gaussian distributed, with the important property that its average power is proportional to the average signal power.
On this basis, the transmitter and the receiver distortion noises are modeled as
\begin{align} \label{eq:etat-distribution}
\boldsymbol{\eta}_t &\sim \mathcal{CN}\left(\mathbf{0}, \delta_t^2  \diag(q_1,\ldots,q_{N_t}) \right) \\
\boldsymbol{\eta}_r &\sim \mathcal{CN}\left(\mathbf{0}, \delta_r^2  \mathrm{tr}( \mathbf{Q}) \mathbf{I}_{N_r} \right) \label{eq:etar-distribution}
\end{align}
where $q_1,\ldots,q_{N_t}$ are the diagonal elements of the signal covariance matrix $\mathbf{Q}$. This means that the transmitter distortion power at the $n$th transmit antenna is proportional to the signal power $q_n$ applied on the same antenna, while the receiver distortion power at the $m$th receive antenna is proportional to the average signal power $\mathrm{tr}( \mathbf{Q})$ received over the $m$th row of the channel matrix $\mathbf{H}$.\footnote{In practice, the receiver will amplify and filter the received signal ${\mathbf{y}}$ in several steps during reception. The receiver distortion noise $\boldsymbol{\eta}_r$ is the aggregation of these steps which occur at different amplification levels. Without loss of generality, \eqref{eq:etar-distribution} represents the aggregate receiver distortion using the amplification at the time when the signal first reaches the receive antenna.} This model assumes sufficient decoupling between the transmit and the receive RF chains, such that the corresponding impairments are statistically independent across the antennas \cite{studer2010residual}. 

The proportionality parameters $\delta_t$ and $\delta_r$ characterize the level of residual impairments in the transmitter and receiver, respectively. Note that $\delta_t$ and $\delta_r$ are inherently connected to the error-vector magnitude (EVM) metric \cite{holma2011LTE}, which is commonly used to quantify the mismatch between the intended signal and the actual signal in RF transceivers. In our notation, the total EVM in the transmitter hardware is given by
\begin{equation}
\mathrm{EVM} \triangleq \sqrt{\frac{\mathbb{E}_{\boldsymbol\eta_t}[ \left\| \boldsymbol\eta_t \right\|^2 ]}{\mathbb{E}_{\mathbf s}[ \left\| \mathbf s\right\|^2 ]}} = \delta_t.
\end{equation}
In practical applications, such as long term evolution (LTE), the EVM requirements are in the range $\delta_t \in [0.08, 0.175]$ \cite[Sec.~14.3.4]{holma2011LTE}. Note that (\ref{eq:impairmodel}) reduces to the canonical model (\ref{eq:canonical}) for $\delta_t = \delta_r = 0$, which indicates ideal hardware on both sides.

\section{Ergodic Channel Capacity Analysis}

Based on the residual impairment model in Section \ref{sec:signalandsystems}, we now derive an expression for the ergodic capacity for any number of antennas and arbitrary SNR.

\begin{definition}
The SNR is denoted by $\rho$. We will also use the SNR to represent the effective signal power $\rho = \mathrm{tr}( \mathbf{Q}  )$, since the channel gain and receiver noise power are normalized in this paper.
\end{definition}

\begin{remark}
An increase in SNR can be achieved by increasing the transmit power or improving the channel conditions (i.e., by decreasing the propagation loss). If the transmit power is increased such that we move far outside the dynamic range of the power amplifier, then the EVM will increase and thus $\delta_t$ should also be increased \cite{studer2011residual2} and advanced dynamic power adaptation is required \cite{Emil2012imp}. For the sake of clarity and brevity in interpretation, we will keep the analysis clean by assuming that any change in SNR is achieved by a change in the propagation loss, while the transmit power is fixed.
\end{remark}

We define $q \triangleq \mathrm {min}({N_t,N_r})$ and $p \triangleq \mathrm {max}({N_t,N_r})$ and have the following lemma.

\begin{lemma} \label{lemma_capacity_expression}
The ergodic channel capacity of \eqref{eq:impairmodel} with i.i.d.~Rayleigh fading and under the constraint $\mathrm{tr}( \mathbf{Q}  ) \leq \rho$ is
\begin{equation} \label{eq:ergodiccapacity}
C = {{\mathbb E}_{\mathbf H}} \left[ \log_2 \det \left( \mathbf{I}_{N_r} + \frac{\rho}{N_t} \mathbf{H} \mathbf{H}^H \boldsymbol{\Phi}^{-1} \right) \right]
\end{equation}
where $\boldsymbol{\Phi} \triangleq \left( \frac{\rho}{N_t} \delta_t^2 \mathbf{H} \mathbf{H}^H + (\delta_r^2 \rho+ 1) \mathbf{I}_{N_r} \right)$. The capacity is achieved by $\mathbf{s} \sim \mathcal{CN}(\mathbf{0}, \frac{\rho}{N_t} \mathbf{I}_{N_t})$, giving the distortion noise distributions
\begin{align}
\boldsymbol{\eta}_t &\sim \mathcal{CN}\left(\mathbf{0}, \delta_t^2  \frac{\rho}{N_t} \mathbf{I}_{N_t} \right) \\
\boldsymbol{\eta}_r &\sim \mathcal{CN}\left(\mathbf{0}, \delta_r^2  \rho \mathbf{I}_{N_r} \right).
\end{align}
\end{lemma}
\begin{IEEEproof}
For any channel realization $\mathbf{H}$ and signal covariance matrix $\mathbf{Q}$, \eqref{eq:impairmodel} can be seen as an instance of the canonical model \eqref{eq:canonical} with noise covariance
\begin{equation}
\boldsymbol{\Phi} = \left( \delta_t^2 \mathbf{H} \diag(q_1,\ldots,q_{N_t}) \mathbf{H}^H \!+\! (\delta_r^2  \mathrm{tr}( \mathbf{Q})+ 1) \mathbf{I}_{N_r} \right).
\end{equation}
Thus, the sufficiency of using a Gaussian distribution on $\mathbf{s}$ follows from \cite{telatar1999capacity} and the ergodic capacity becomes
\begin{equation}
C = \max_{\mathbf{Q}: \, \mathrm{tr}( \mathbf{Q}  ) \leq \rho} {{\mathbb E}_{\mathbf H}} \bigg[{ \log_2 \det \left( \mathbf{I}_{N_r} + \mathbf{H} \mathbf{Q} \mathbf{H}^H \boldsymbol{\Phi}^{-1} \right) }\bigg].
\end{equation}
Finally, the optimality of the signal covariance matrix $\mathbf{Q} = \frac{\rho}{N_t} \mathbf{I}_{N_t}$ is a simple consequence of \cite[Corollary 1]{Emil2013imp}.
\end{IEEEproof}

This lemma shows that the ergodic capacity with transceiver hardware impairments has a similar structure as for the canonical model in \eqref{eq:canonical}. In the remainder of this paper we will, however, show that it behaves fundamentally different in many cases and regimes of practical relevance.

\subsection{Exact SNR Analysis}

We now derive a closed-form expression for the ergodic capacity in Lemma \ref{lemma_capacity_expression}. To this end, we first define the instantaneous MIMO channel correlation matrix as
\begin{equation} \label{eq:W-definition}
\mathbf W \triangleq \begin{cases}
\mathbf H\mathbf H^H, & \textrm{if $N_r\leq N_t$},\\
\mathbf H^H\mathbf H, & \textrm{if $N_r > N_t$},
\end{cases}
\end{equation}
since it will be often used in our manipulations. We begin our discussion with the following proposition.

\begin{proposition} \label{prop:ergodic-capacity}
For i.i.d. Rayleigh fading channels, the ergodic capacity in \eqref{eq:ergodiccapacity}, under the proposed residual impairment model, is
\begin{equation}
\begin{split}
C_{N_r\times N_t} &= \frac{q}{{\ln (2)}}{K}\sum\limits_{n = 1}^q {\sum\limits_{m = 1}^q {{{\left( { - 1} \right)}^{n + m}} \det \left( \boldsymbol\Omega  \right)} }  \Gamma \left( {t + 1} \right) \\
&\times\sum\limits_{k = 1}^{t + 1}{\left( {e^{\frac{1}{f}}}{ E}_{t+2-k}\left( \frac{1}{f}\right)  - {e^{\frac{1}{g}}} E_{t+2-k}\left( \frac{1}{g}\right) \right)}\label{eq:analcapacity}
\end{split}
\end{equation}
where we define $t \triangleq n+m+p-q-2$, $f \triangleq \frac{\rho\left(\delta_t^2 +1 \right)}{N_t\left(\rho\delta_r^2+1\right)}$ and $g \triangleq \frac{\rho\delta_t^2}{N_t(\rho\delta_r^2 + 1)}$, while ${K} = \left[ \prod_{i=1}^{q}(p-i)!\prod_{j=1}^{q}(q-j)!  \right]^{-1}$ is a normalization constant. Moreover, $\Gamma(z)$  denotes the Gamma function \cite[Eq.~(8.310.1)]{gradshteyn2007a} and $E_n(z) = \int_1^\infty t^{-n}e^{-zt}dt$  is the exponential integral function \cite[Eq.~(8.211.1)]{gradshteyn2007a}. Finally, $\boldsymbol\Omega$ is a $(q-1)\times(q-1)$ matrix whose $(i,j)$-th element is given by
\begin{equation}
\boldsymbol\Omega_{i,j}=\left(\alpha_{i,j}^{(n)(m)} + p - q\right)! \, q^{-\frac{1}{q-1}}\notag
\end{equation}
where
\begin{equation}
\alpha_{i,j}^{(n)(m)} \triangleq \begin{cases}
i+j-2, & \textrm{if $i<n$ and $j<m$}\\
i+j, & \textrm{if $i\ge n$ and $j \ge m$}\\
i+j-1, & \textrm{otherwise.}
\end{cases}
\end{equation}

\begin{IEEEproof}
Using the notation in \eqref{eq:W-definition}, the ergodic capacity in \eqref{eq:ergodiccapacity} can be expressed as
\begin{align}
\mathit{C}_{N_r\times N_t} = {{\mathbb E}_{\mathbf H}} \left[ {\log _2}\det \left( {\mathbf{I}_q + \frac{\rho}{N_t}\mathbf{W}{\boldsymbol\Phi^{-1}}} \right)\right]
\label{eq:detcapacity}
\end{align}
where $\boldsymbol{\Phi} = \left( \frac{\rho}{N_t} \delta_t^2 \mathbf{W} + (\delta_r^2 \rho+ 1) \mathbf{I}_{q} \right)$. Note that $\mathbf W$ is a $q\times q$ random, non-negative definite matrix following the complex Wishart distribution. Thus, it has real non-negative eigenvalues and the probability density function (PDF) of its unordered eigenvalue, $\lambda$, is found in \cite[Eq.~(38)]{zanella2009marginal} to be
\begin{equation}
p_\lambda(\lambda) = K\sum\limits_{n = 1}^{q}\sum\limits_{m = 1}^{q} \frac{(-1)^{m+n} \lambda^{n+m+p-q-2}}{ e^{\lambda}} \det \left({\boldsymbol\Omega}\right).
\end{equation}

By exploiting the eigenvalue properties, we can now alternatively express the capacity in (\ref{eq:detcapacity}) as
\begin{align}
&C_{N_r\times N_t} = {{\mathbb E}_{\mathbf H}}\left[ {\sum\limits_{i = 1}^q {{{\log }_2}\left( {1 + \frac{{\frac{{\rho}}{{N_t}}{\lambda _i}}}{{\frac{{\rho\delta_t^2}}{{{N_t}}}{\lambda _i} + \rho\delta_r^2 + 1}}} \right)} } \right] \label{C_alter}  \\&=q\int\limits_0^\infty  {{{\log }_2}\left( {1 + \frac{{\frac{\rho}{N_t}\lambda }}{{\frac{\rho\delta_t^2}{N_t}\lambda  + \rho
\delta_r^2+1}}} \right){p_\lambda }} \left( \lambda  \right)d\lambda \label{eq:CapIntegral}\\
&= q\left( \int\limits_0^\infty  {{{\log }_2}\left( {\left( {\frac{{\rho}}{N_t} + \frac{{\rho\delta_t^2}}{{{N_t}}}} \right)\lambda  + \rho\delta_r^2 + 1} \right) {p_\lambda }} \left( \lambda  \right)d\lambda \right)\notag\\ & \quad - q\left(\int\limits_0^\infty  {{{\log }_2}\left({\frac{{\rho\delta_t^2}}{{{N_t}}} \lambda  + \rho\delta_r^2 + 1} \right)} {p_\lambda }\left( \lambda  \right)d\lambda \right)\label{eq:int}
\end{align}
where $\lambda_i$ represents the $i$-th ordered eigenvalue of $\mathbf W$.

The integrals in (\ref{eq:int}) can be evaluated using the following integral identity \cite[Eq.~(40)]{kang2006rician}
\begin{align}
\int\limits_0^\infty\ln\left(1+ay\right)y^{n-1}e^{-cy}dy = \Gamma(n)e^{\frac{c}{a}}\sum\limits_{k = 1}^{n}\frac{\Gamma\left(-n+k, \frac{c}{a}\right)}{c^ka^{n-k}}
\end{align}
and the fact that $E_n(z) = z^{n-1}\Gamma(1-n,z)$, where $\Gamma(s,x)=\int_x^\infty t^{s-1}e^{-t}dt$ is the upper incomplete gamma function \cite[Eq.~(8.350.2)]{gradshteyn2007a}. The expression in \eqref{eq:analcapacity} then follows after some simple algebraic manipulations.
\end{IEEEproof}
\end{proposition}


%
%

\begin{figure}[ht]
\begin{centering}
\includegraphics [keepaspectratio,width=\columnwidth]{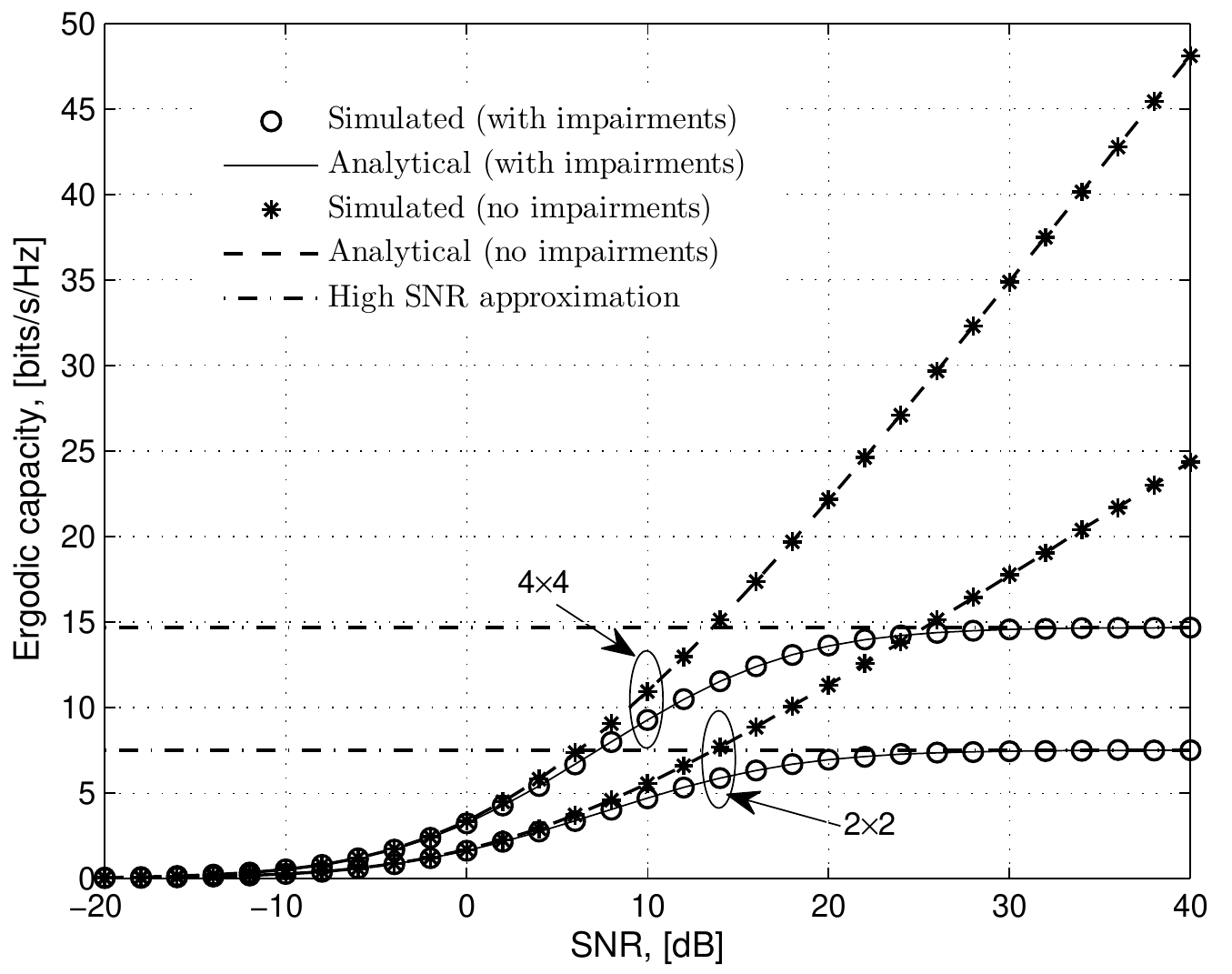}
\caption{Simulated and analytical ergodic MIMO capacity in i.i.d. Rayleigh fading channels with residual hardware impairments ($\delta_t = \delta_r = 0.15$) and without hardware impairments ($\delta_t = \delta_r = 0$).}\label{fig:Sim_Analytical}
\end{centering}
\end{figure}

Figure \ref{fig:Sim_Analytical} illustrates the ergodic capacity for different hardware conditions and antenna configurations. In all cases, the results demonstrate an excellent agreement between analytical results and Monte-Carlo simulations. Furthermore, for both $2\times2$ and $4\times4$ configurations, it is clear that hardware impairments will cause severe degradation on the ergodic capacity, compared with the ideal case studied by Telatar and Foschini \cite{telatar1999capacity,foschini1998a}. Observe that the capacity gap between the ideal system and the impaired system gets larger with the SNR. It is also noteworthy that for high SNR values, the ergodic capacity saturates and thus exhibits a finite capacity ceiling that cannot be crossed regardless of the SNR value. The reason for this effect is that the distortion noise power on both sides grows linearly and unboundedly with the transmit power. This confirms that hardware impairments fundamentally limit the performance of high-capacity systems, as quantified by the following corollary.
\begin{corollary} \label{cor:asymptotic-capacity}
Asymptotically as $\rho \rightarrow \infty$, the ergodic capacity in (\ref{eq:ergodiccapacity}) approaches the finite limit
\begin{equation}
\begin{split}
{C_\mathrm{limit} }\!& = \frac{q}{{\ln (2)}}{K}\sum\limits_{n = 1}^q {\sum\limits_{m = 1}^q {{{\left( { - 1} \right)}^{n + m}} \det \left( \boldsymbol\Omega  \right)} }  \Gamma \left( {t + 1} \right) \\
&\times\sum\limits_{k = 1}^{t + 1}{\left( {e^{\frac{1}{\hat{f}}}}{ E}_{t+2-k}\left( \frac{1}{\hat{f}}\right) - {e^{\frac{1}{\hat{g}}}} E_{t+2-k}\left( \frac{1}{\hat{g}}\right) \right)} \label{eq:CapLimit}
\end{split}
\end{equation}
where $\hat{f} \triangleq \frac{1+\delta_t^2}{N_t\delta_r^2}$ and $\hat{g} \triangleq \frac{\delta_t^2}{N_t\delta_r^2}$.
\vspace{0.1cm}
\begin{IEEEproof}
The asymptotic capacity is defined as
\begin{align}
{C_\mathrm{limit} }\!&= \!\mathop {\lim }\limits_{\rho \to \infty } \!{\mathbb E}_\mathbf{H}\!\left[ {\sum\limits_{i = 1}^q {{{\log }_2}\left( {1 + \frac{{\frac{{\rho}}{{N_t}}{\lambda _i}}}{{\frac{{\rho\delta_t^2}}{{{N_t}}}{\lambda_i} + \rho\delta_r^2 + 1}}} \right)} } \right]\label{eq:CapLimit1}\\
  &= q\int_0^\infty{{\log }_2}\left( {1 + \frac{\lambda}{{\delta_t^2\lambda+N_t\delta_r^2}}} \right)p(\lambda)d(\lambda). \label{eq:CapLimit2}
\end{align}
From (\ref{eq:CapLimit1}) to (\ref{eq:CapLimit2}) we have changed the order of expectation and limit, since according to Jensen's inequality,
the term inside the expectation is upper bounded by an integrable function, hence the dominated convergence theorem \cite{bartle1995integration} holds. The final expression \eqref{eq:CapLimit} is obtained as in Proposition \ref{prop:ergodic-capacity}.
\end{IEEEproof}
\end{corollary}

As expected, \eqref{eq:CapLimit} is a deterministic constant independent of the SNR. Observe that Corollary \ref{cor:asymptotic-capacity} extends the asymptotic results in \cite{studer2010residual,Emil2013imp} that only considered aggregate hardware impairments at the receiver.



\vspace{0.05cm}
\subsection{Low-SNR Analysis}
In the low-SNR regime, the capacity is well-approximated as \cite{lozano2003lowSNR}
\begin{align}
C \left(\frac{E_b}{N_0}\right) \approx S_0\log_2\left(\frac{\frac{E_b}{N_0}}{\frac{E_b}{N_0}_{\mathrm{min}}}\right)
\end{align}
where $\frac{E_b}{N_0}_\mathrm{min}$ and $S_0$ represent the \emph{minimum normalized energy per information bit required to convey any positive rate reliably} and the \emph{wideband slope}, respectively.

As shown in \cite{lozano2003lowSNR}, $S_0$ and $\frac{E_b}{N_0}_{\mathrm{min}}$ can be calculated from the first and second derivatives of $C(\rho)$ at $\rho = 0$ through
\begin{align}
\frac{E_b}{N_0}_{\mathrm{min}} \triangleq \mathop{\lim}\limits_{\rho \to 0} \frac{\rho}{C(\rho)} = \frac{1}{\dot{C}(0)}\label{eq:EbNo_def}
\end{align}
and
\begin{align}
S_0 \triangleq -\frac{2 \ln (2) \left[\dot{C}(0)\right]^2}{\ddot{C}(0)}\label{eq:So_def}.
\end{align}

\begin{proposition}
For i.i.d. Rayleigh fading channels, the minimum energy per information bit and the wideband slope, under the proposed residual impairment model, are respectively given by
\begin{align}
\frac{E_b}{N_0}_{\mathrm{min}} &= \frac{\ln (2) }{N_r} \label{eq:EbNo_result}
\\S_0 &= \frac{2N_tN_r}{(2\delta_t^2+1)(N_t+N_r)+2\delta_r^2N_t}\label{eq:So_result}.
\end{align}

\end{proposition}

\begin{IEEEproof}
Substituting (\ref{eq:CapIntegral}) into (\ref{eq:EbNo_def}) and taking the first derivative with respect to $\rho$, we get
\begin{align}
\dot{C}(0)&=q\int_0^\infty\!{\left[{{\log }_2}\left( \left.{1\!+\!\frac{{\frac{\rho}{N_t}\lambda }}{{\frac{\rho\delta_t^2}{N_t}\lambda\!+\!\rho\delta_r^2\!+\!1}}} \right)'\right|_{\rho=0}\right]{p_\lambda }} \left( \lambda  \right)d\lambda\notag\\
&=\frac{q}{N_t \ln(2)}\int_0^\infty\lambda p(\lambda) d\lambda\label{eq:C1st_1}\\
&=\frac{\mathbb E\left[ \mathrm{tr}\left(\mathbf W\right) \right]}{N_t \ln (2)} = \frac{N_r}{\ln (2)}\label{eq:C1st_2}
\end{align}
where from (\ref{eq:C1st_1}) to (\ref{eq:C1st_2}) we have used the fact that \cite[Lemma~
4]{lozano2003lowSNR}
\begin{equation}
q\int_0^\infty \lambda p(\lambda)d\lambda = q\mathbb E[\lambda] = \mathbb E\left[\mathrm{tr}\left(\mathbf W\right)\right] = N_rN_t.
\end{equation}
Similarly, we can find the second derivative as follows
\begin{align}
\ddot{C}(0) &=q\int_0^\infty\!{\left[{{\log }_2}\left( \left.{\!1\!+\!\frac{{\frac{\rho}{N_t}\lambda }}{{\frac{\rho\delta_t^2}{N_t}\lambda\! +\! \rho\delta_r^2\!+\!1}}} \right)''\right|_{\rho=0}\right]{p_\lambda }} \left( \lambda  \right)d\lambda\notag\\
&=-\frac{q(2\delta_t^2+1)}{\ln (2) N_t^2}\!\int_0^\infty\!\lambda^2p(\lambda)d\lambda \!-\! \frac{2q\delta_r^2}{\ln(2) N_t}\!\int_0^\infty\!\lambda p(\lambda)d\lambda \label{eq:C2nd_1}\!\\
&=-\frac{q(2\delta_t^2+1)}{\ln(2) N_t^2}\mathbb E\big[\mathrm{tr}\left(\mathbf W^2\right)\big] - \frac{2q\delta_r^2}{\ln(2) N_t}\mathbb E\big[\mathrm{tr}\left(\mathbf W\right)\big]\label{eq:C2nd_2}\\
&=-\frac{N_r}{\ln (2) }\left( \frac{(2\delta_t^2+1)(N_t+N_r)}{N_t} + 2\delta_r^2 \right)\label{eq:C2nd_3}
\end{align}
where from (\ref{eq:C2nd_1}) to (\ref{eq:C2nd_3}) we use the fact that \cite[Lemma~4]{lozano2003lowSNR}
\begin{equation}
q\int_0^\infty \lambda^2 p(\lambda)d\lambda\!= \!q\mathbb E[\lambda^2]\!= \!\mathbb E\left[\mathrm{tr}\left(\mathbf W^2\right)\right]\!=\! N_rN_t\left(N_r+N_t\right).
\end{equation}
Combining (\ref{eq:C1st_2}), (\ref{eq:C2nd_3}) with the definitions in (\ref{eq:EbNo_def}) and (\ref{eq:So_def}) we can obtain the results in (\ref{eq:EbNo_result}), (\ref{eq:So_result}).
\end{IEEEproof}

\begin{figure}[t]
\begin{centering}
\includegraphics [keepaspectratio,width=\columnwidth]{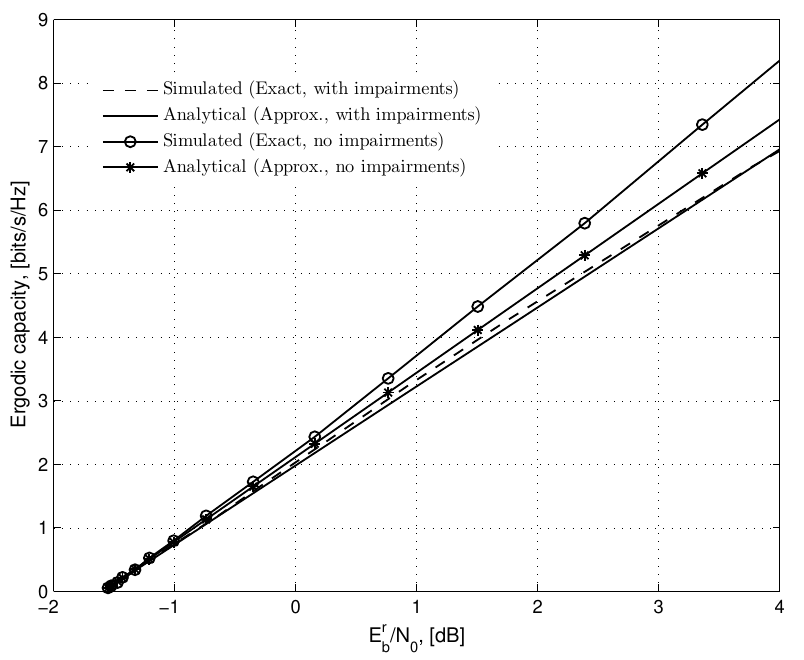} \vskip-2mm
\caption{Simulated and analytical low-SNR ergodic capacity of a $4\times4$ MIMO system in i.i.d. Rayleigh fading channels with impairments ($\delta_t = \delta_r = 0.15$) and without impairments ($\delta_t = \delta_r = 0$).}\label{fig:LowSNR} \vskip-3mm
\end{centering}
\end{figure}

Figure~\ref{fig:LowSNR} depicts the ergodic capacity of a $4\times4$ MIMO system under different hardware conditions. Note that $\frac{E_b}{N_0}_\mathrm{min}$ is the intersection of the curves with the horizontal axis. We see in both cases that the analytical results (linear approximation) and the numerical results have very good agreement across a wide SNR range. Interestingly, for both ideal and impaired systems, $\frac{E_b}{N_0}_\mathrm{min}$ remains the same. The impact of transceiver impairments is seen only via the wideband slope $S_0$; observe that the slope of capacity curve decreases when impairments are considered. This implies that hardware impairments have only a second-order impact on the capacity in the low-SNR regime. From the expression in \eqref{eq:So_result}, we notice that the transmitter impairments have a more influential impact on $S_0$ than the receiver impairments, since $\delta_t^2$ is multiplied with a larger number. This reveals that transmitter impairments are more influential in the low-SNR regime.

\vspace{-0.2cm}

\subsection{Large-Scale MIMO Analysis}

In this section, we analyze the asymptotic behavior of the ergodic capacity when the number of antennas is large. Specifically, our discussion includes three cases:
\begin{enumerate}[i)]
  \item large $N_t$ and fixed $N_r$;\label{case1}
  \item large $N_r$ and fixed $N_t$;
  \item large $N_t$ and $N_r$, with a fixed finite ratio $\beta = \frac{N_r}{N_t}>0$.
\end{enumerate}

\begin{figure}[t]
\begin{centering}
\includegraphics [keepaspectratio,width=\columnwidth]{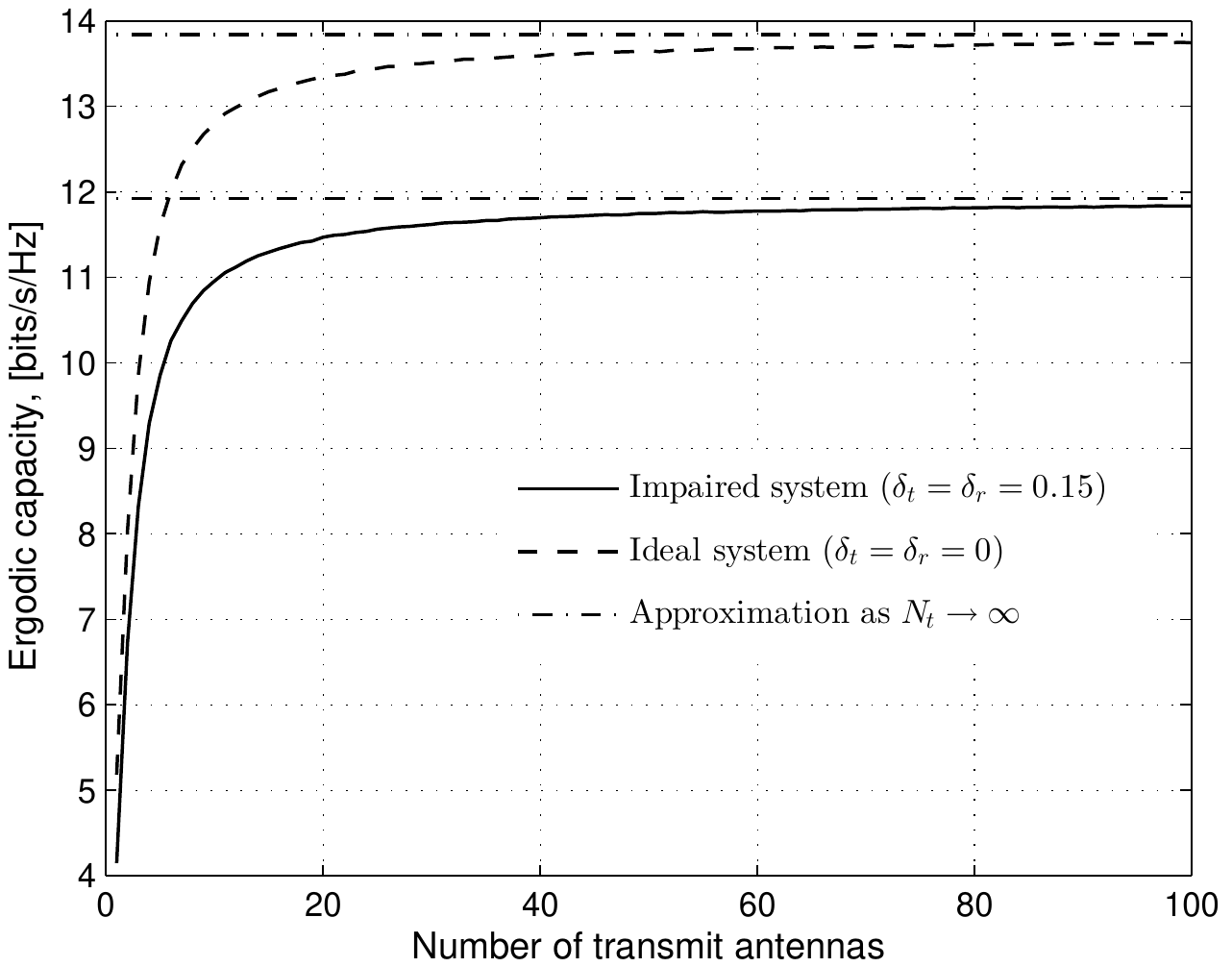} \vskip-2mm
\caption{Ergodic MIMO capacity when the numbers of receive antennas is fixed and the number of transmit antennas increases ($N_r = 4, \rho = 10\mathrm{dB}$).}\label{fig:Nrfix}
\end{centering}
\end{figure}

For case i), recall that the $N_r \times N_r$ matrix $\frac{1}{N_t}\mathbf H\mathbf H^H$ converges almost surely to $\mathbf I_{N_r}$ almost surely as $N_t \rightarrow \infty$ \cite{bertrand2004hardening}, thus \eqref{eq:ergodiccapacity} becomes
\begin{equation}
C_{\infty, N_r} = N_r\log_2\left( 1 + \frac{\rho}{\rho\delta_t^2 + \rho\delta_r^2 + 1}  \right).\label{eq:Ntinf}
\end{equation}
This shows that having a large number of transmit antennas makes the capacity converge to a finite deterministic value, which is characterized by the level of transceiver impairments and the number of receive antennas. From (\ref{eq:Ntinf}), we notice that for fixed SNR values, the capacity increases linearly with the number of receive antennas; however, for fixed antenna setups, if we only increase SNR, the capacity behaves similar to the finite-dimensional MIMO case in Corollary \ref{cor:asymptotic-capacity}; that is, it saturates in the high-SNR regime. Figure \ref{fig:Nrfix} shows the behavior of the ergodic capacity as we increase the number of transmit antennas. When $N_t$ is small, the gap between the two systems is small; as $N_t$ increases the gap gets larger, and finally when $N_t$ is sufficiently large the gap converges to a constant which is determined by the SNR and the level of impairments. We can interpret this gap as a SNR penalty due to the transceiver hardware impairments. From Fig.~\ref{fig:Nrfix}, we can see that, for both the impaired system and the ideal system, the ergodic capacity converges to the finite ceiling given by \eqref{eq:Ntinf} as $N_t$ grows large.
\vspace{-0.3cm}

\begin{figure}[t]
\begin{centering}
\includegraphics [keepaspectratio,width=\columnwidth]{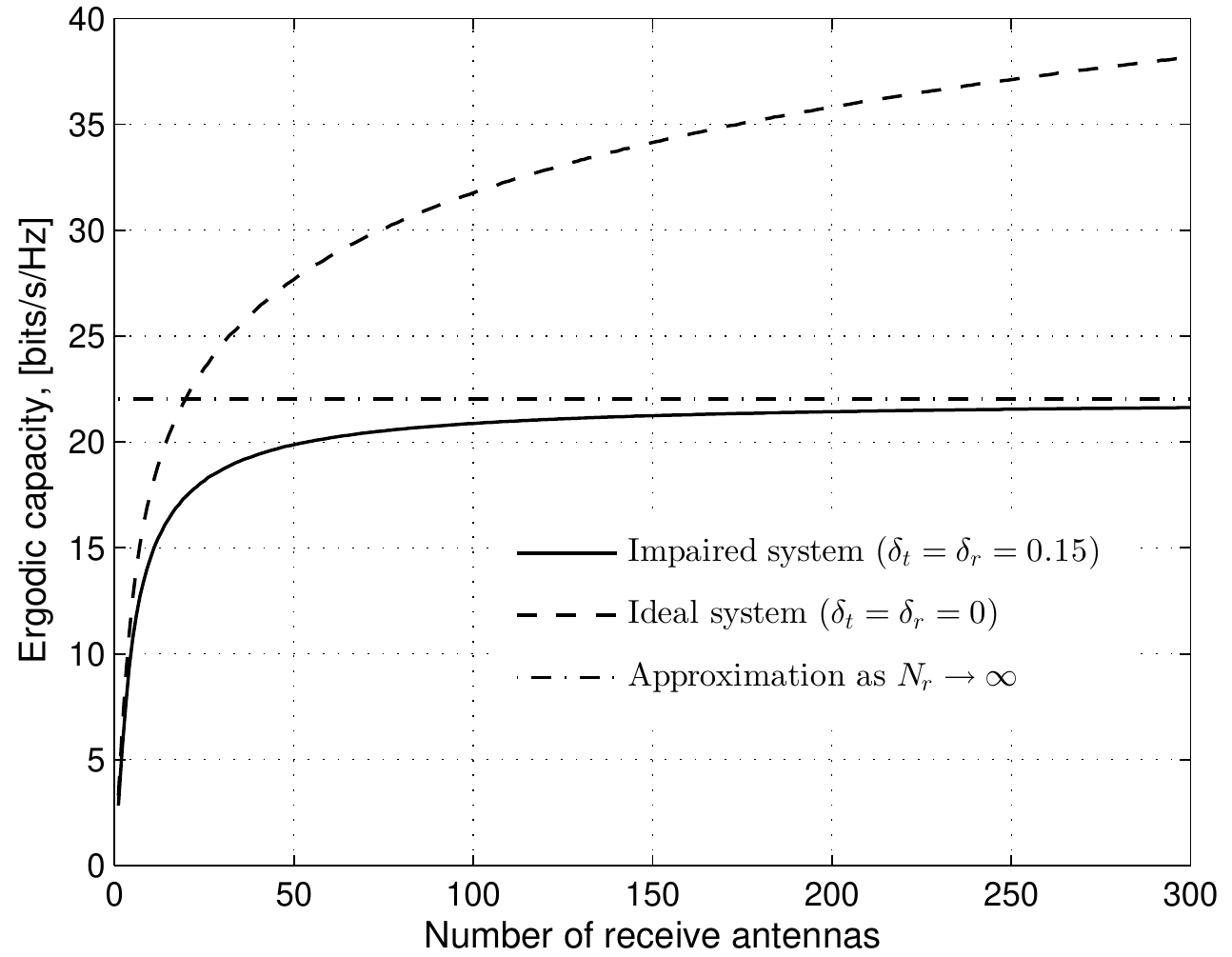}
\caption{Ergodic MIMO capacity when the number of transmit antennas is fixed and the number of receive antennas increases ($N_t = 4, \rho = 10\mathrm{dB}$).}\label{fig:Ntfix}
\end{centering}
\end{figure}

For case ii), analogously to the previous case, we use the property that $\frac{1}{N_t}\mathbf H^H\mathbf H = \frac{N_r}{N_t}\times\frac{1}{N_r} \mathbf H^H\mathbf H$ and the $N_t\times N_t$ matrix $\frac{1}{N_r}\mathbf H^H\mathbf H$ converges to $\mathbf I_{N_t}$ as $N_r \rightarrow \infty$ \cite{bertrand2004hardening}. Consequently, \eqref{eq:ergodiccapacity} becomes
\begin{align}
C_{N_t, \infty} &= \mathop {\lim }\limits_{N_r \to \infty } N_t\log_2\left( 1 +  \frac{\rho\frac{N_r}{N_t}}{\rho\delta_t^2\frac{N_r}{N_t}+\rho\delta_r^2+1} \right) \notag\\
&= N_t\log_2\left( 1+\frac{1}{\delta_t^2} \right).\label{eq:Nrinf}
\end{align}
Observe that this capacity ceiling is characterized only by the number of transmit antennas and the level of transmitter impairments, while the receiver impairments have no impact on the system performance, similar to \cite[Theorem 1]{Emil2013imp}, where receiver impairments were ignored all along.
The behavior in case ii) is quite different from case \ref{case1}), where both the transmitter and the receiver impairments affect the system performance. This result indicates a very important implication for system design: if a large-scale MIMO system adopts the antenna configuration with $N_r \gg N_t$, it is of pivotal importance to build high-quality transmitter hardware. Figure \ref{fig:Ntfix} compares the ergodic capacity of the impaired system and the ideal system as $N_r$ increases. We find that, for the impaired system, the ergodic capacity converges to the finite deterministic ceiling which is given in \eqref{eq:Nrinf}. However, for the ideal system, the capacity increases logarithmically with $N_r$ (this coincides with the result in \cite[Eq.~(6)]{bertrand2004hardening}). We conclude from this observation that, with the existence of residual transmitter hardware impairments, we cannot benefit from the array gain by increasing the number of receive antennas as in the ideal hardware case.

\begin{figure}[ht]
\begin{centering}
\includegraphics [keepaspectratio,width=\columnwidth]{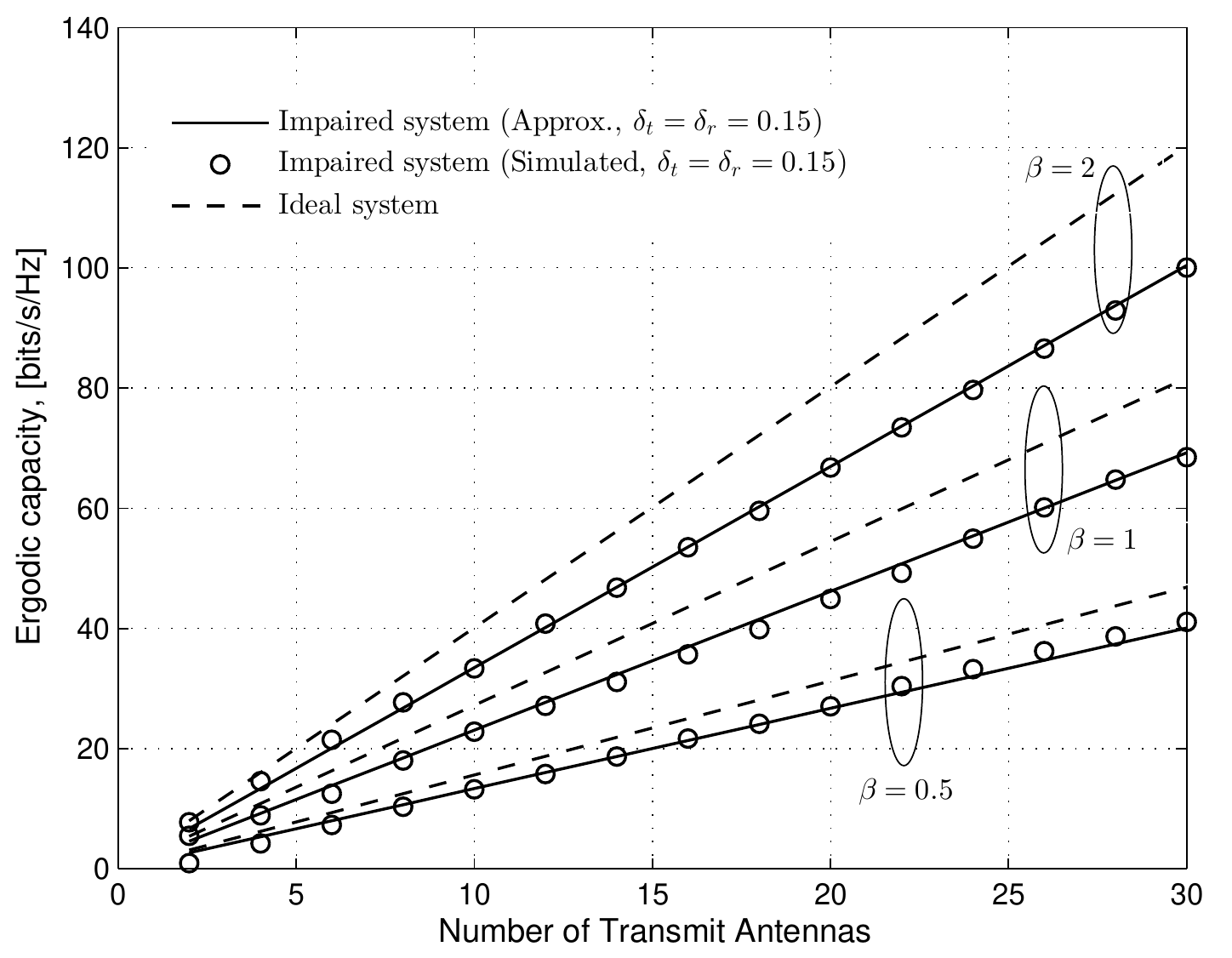}
\caption{Ergodic MIMO capacity when the number of transmit and receive antennas both increase with a fixed ratio ($\beta=\frac{N_r}{N_t}, \rho = 10\mathrm{dB}$). }\label{fig:Large_Nr_Nt}
\end{centering}
\end{figure}

For case iii), we first note that the ergodic capacity in \eqref{eq:ergodiccapacity} can be expressed as the difference
\begin{equation}
\begin{split}
C &= {{\mathbb E}_{\mathbf H}} \left[ \log_2 \det \left( (\delta_r^2 \rho+ 1) \mathbf{I}_{N_r} + \frac{\rho(1+\delta_t^2)}{N_t} \mathbf{H} \mathbf{H}^H  \right) \right] \\ &- {{\mathbb E}_{\mathbf H}} \left[\log_2 \det \left( (\delta_r^2 \rho+ 1) \mathbf{I}_{N_r} + \frac{\rho \delta_t^2}{N_t} \mathbf{H} \mathbf{H}^H  \right) \right]
\end{split}
\end{equation}
where each term represents a classical ergodic capacity expression. The asymptotic behavior as $N_t$ and $N_r$ grow large
with a finite ratio $\beta = \frac{N_r}{N_t}>0$ can therefore be obtained by analyzing each term separately. More precisely, we apply
an asymptotically tight approximation from \cite[Chapter 13.2.2]{Couillet2011a}, which shows that
\vspace{0.15cm}
\begin{equation} \label{eq:deterministic_equivalent}
\begin{split}
C &- N_r \Bigg[ \!
\log_2 \! \!  \left( \!\frac{1+\rho \delta_r^2 + \frac{\rho(1+\delta_t^2)}{1+\beta \rho_1}
}{1+\rho \delta_r^2 + \frac{\rho\delta_t^2}{1+\beta \rho_2}} \! \right) \!\! +\! \beta^{-1} \log_2 \! \left( \frac{1 + \beta \rho_1}
{1 + \beta \rho_2} \right) \\ & + \log_2(e) \left( \frac{\rho_1 (1+\rho \delta_r^2) }{\rho(1+\delta_t^2)} - \frac{\rho_2 (1+\rho \delta_r^2) }{\rho\delta_t^2} \right)
 \Bigg] = \mathcal{O} \left(\frac{1}{N_t} \right)
 \end{split}
\end{equation}
\vspace{0.15cm}
where $C$ is the true ergodic capacity, $L \triangleq 1 - \beta^{-1}$, and
\begin{align}
\rho_1 & \triangleq \frac{1}{2} \Bigg( \frac{\rho(1+\delta_t^2) L}{1+\rho \delta_r^2} - \beta^{-1} \notag\\ & + \sqrt{ \Big( \frac{\rho(1+\delta_t^2) L}{1+\rho \delta_r^2} - \beta^{-1} \Big)^2 + \frac{4  \rho (1+\delta_t^2)}{ \beta (1+\rho \delta_r^2)}  } \Bigg)  \\
\rho_2 & \triangleq \frac{1}{2} \Bigg( \frac{\rho\delta_t^2 L}{1+\rho\delta_r^2} - \beta^{-1} \notag\\ & + \sqrt{ \Big(\frac{\rho\delta_t^2 L}{1+\rho \delta_r^2} - \beta^{-1} \Big)^2 + \frac{4 \rho \delta_t^2}{\beta (1+\rho \delta_r^2)}  } \Bigg).
\end{align}
Since the difference between the true and approximate ergodic capacity in \eqref{eq:deterministic_equivalent} behaves as $\mathcal{O}\left(\frac{1}{N_t}\right)$, the approximation error vanishes asymptotically. Observe that the expression inside the square brackets in \eqref{eq:deterministic_equivalent} is strictly positive and only depends on the fixed ratio $\beta=\frac{N_r}{N_t}$ and not on the individual values on $N_t$ and $N_r$. When the number of antennas is scaled with the fixed ratio $\beta$, then the approximated ergodic capacity in \eqref{eq:deterministic_equivalent} grows linearly with $N_r$ and without bound. This is the same scaling behavior as with ideal hardware, which means that transceiver impairments will only inflict a small/bounded offset in the ergodic capacity in this case. It is important to note that since the SNR is fixed, the effective SNR per element in $\mathbf{s}$ will reduce as $\frac{\rho}{N_t}$, but this reduction is counteracted by the large array gain achieved at the receiver when $N_r$ also grows large. Another important explanation for this phenomenon is that by deploying a large number of antennas on both sides, we can create several spatially parallel subchannels. The total transmit power is then allocated to these subchannels such that each subchannel only has a small portion of the total transmit power, which makes the effective SNR on each stream very low. As we observed in the low-SNR analysis, transceiver impairments do not have significant impact on the capacity for low SNR values. Consequently, the ``capacity ceiling'' disappears. Figure \ref{fig:Large_Nr_Nt} illustrates the ergodic capacity of this scenario for three different values of $\beta$. Apart from the previous observations, this figure also demonstrates that for different $\beta$, the relative capacity gaps, $\frac{C_\mathrm{imp} - C_\mathrm{ideal}}{C_\mathrm{ideal}}$, between the impaired systems and the ideal systems, are nearly the same. This indicates that residual transceiver impairments has a  relative smaller impact on large-dimensional MIMO systems.

\vspace{-0.2cm}
\section{Conclusions}
Residual RF hardware impairments can have a dramatic impact on the capacity of MIMO communication systems, especially on those operating at high SNRs (i.e., high-rate systems). In this paper, we analytically derived an ergodic capacity expression for a MIMO system with residual transceiver impairments, which applies for any finite number of antennas and the entire SNR range. This expression can be very easily evaluated, since it only contains elementary functions. Additionally, we presented analytical capacity expressions in the high-SNR and low-SNR regimes. Finally, we presented results on the ergodic capacity for large-scale MIMO systems with residual transceiver impairments. While the ergodic capacity generally has a finite ceiling due to the transceiver impairments, we found that by increasing both the number of transmit and receive antennas, the ceiling vanishes and the capacity can increase unboundedly at any SNR. As such, large-scale MIMO is one viable solution for mitigating the detrimental impact of residual impairments---at least, if the increasing overhead signaling can be handled properly.

\section*{ACKNOWLEDGMENTS}
We wish to thank Dr.~G.~Durisi for fruitful discussions. The work of X.~Zhang, M.~Matthaiou and M. Coldrey has been supported in part by the Swedish Governmental Agency for Innovation Systems (VINNOVA) within the VINN Excellence Center Chase, and by the Swedish Foundation for Strategic Research. The work of E.~Bj{\"o}rnson was supported by International Postdoc Grant 2012-228 from the Swedish Research Council. This research has been supported by the ERC Starting Grant 305123 MORE (Advanced Mathematical Tools for Complex Network Engineering).

\bibliographystyle{IEEEtran}
\bibliography{IEEEabrv,Reference}

\end{document}